\documentclass[onecolumn]{aastex631}

\usepackage{amsmath}
\usepackage{amssymb}
\usepackage{mathrsfs}
\usepackage[T1]{fontenc}
\usepackage{natbib}
\usepackage{longtable}

\renewcommand{\edit}[2]{#2}

\begin{document}

\title{Do all the quasars and high-excitation radio galaxies (HERGs) in the 3CRR catalog contain a magnetically arrested disk (MAD)?}

\author[0000-0002-7299-4513]{Shuang-Liang Li}
\affiliation{Key Laboratory for Research in Galaxies and Cosmology, Shanghai Astronomical Observatory, Chinese Academy of Sciences, 80 Nandan Road, Shanghai 200030, People's Republic of China}

\author[0000-0002-4521-6281]{Wenwen Zuo}
\affiliation{Key Laboratory for Research in Galaxies and Cosmology, Shanghai Astronomical Observatory, Chinese Academy of Sciences, 80 Nandan Road, Shanghai 200030, People's Republic of China}

\author[0000-0002-2355-3498]{Xinwu Cao}
\affiliation{Institute for Astronomy, School of Physics, Zhejiang University, 866 Yuhangtang Road, Hangzhou 310058, People's Republic of China}

\correspondingauthor{Shuang-Liang Li,Wenwen Zuo, Xinwu Cao}
\email{lisl@shao.ac.cn, wenwenzuo@shao.ac.cn, xwcao@zju.edu.cn}

\begin{abstract}
Based on the magnetization, an accretion disk with large-scale magnetic field can be separated into either standard and normal evolution (SANE) or magnetically arrested disk (MAD), which are difficult to identify from observations. It is still unclear whether all the radio-loud active galactic nuclei (RLAGNs) with a thin disk and strong radio emissions contain a MAD. We investigate this issue by utilizing the 3CRR catalog. We compile a sample of 35 quasars and 14 high-excitation radio galaxies powered by a thin accretion disk. In order to consistently compare with the MAD sample given by \citet{2022A&A...663L...4L}, the optical-UV emissions of our sample are all detected by the Hubble Space Telescope (HST). It is found that the average X-ray luminosity ($L_{\rm X}$) of our sample is about 5.0 times higher than that of radio-quiet AGNs (RQAGNs) with matching optical-UV luminosity ($L_{\rm UV}$), in general accord with the factor of 4.5 times in MAD sample within the uncertainty. The relationship between radio (5~GHz) and X-ray (2 keV) luminosities in the 3CRR sources is also found to be consistent with the MAD sample. Furthermore, the jet efficiencies of 3CRR sources are consistent with those from the GRMHD simulations of MAD. Therefore, we suggest that probably all the quasars and at least a fraction of high-excitation radio galaxies in the 3CRR catalog, and perhaps all the RLAGNs with strong radio emissions contain a MAD.
\end{abstract}

\keywords{accretion, accretion disks -- black hole physics -- magnetohydrodynamics (MHD) -- galaxies: active -- X-rays: galaxies}



\section{Introduction}

\citet{2003PASJ...55L..69N} suggested that with the accumulation of magnetic flux, the magnetic pressure in the inner accretion disk will be comparable or even higher than the gas pressure, leading to the destruction of symmetry in the inner parts of the accretion disk. This kind of disk is called MAD (magnetically arrested disk), where the magneto-rotational instability (MRI) is suppressed and the magnetic Rayleigh–Taylor (RT) instability will be responsible for the accretion of gas (e.g., \citealt{2018MNRAS.478.1837M,2019ApJ...874..168W}). In the general relativistic magnetohydrodynamic (GRMHD) simulations, an advection-domintated accretion flow (ADAF) can be divided into two classes, i.e., the so-called standard and normal evolution (SANE) and MAD according to the magnetization of ADAF (e.g., \citealt{2011MNRAS.418L..79T, 2012MNRAS.423.3083M,2022ApJ...941...30C}). However, it is very difficult to discriminate between MAD and SANE scenarios from observations. Although the jet power produced in MAD is generally larger than that produced in SANE, we cannot distinguish MAD and SANE based on the jet power alone, because it depends not only on the strength of magnetic field but also on the black hole spin and mass accretion rate \citep{1977MNRAS.179..433B}. Additionally, \citet{2019ApJ...887..167X} suggested that the spectral energy distributions (SED) of SANE and MAD in an ADAF are quite similar, though with a higher radiative efficiency for MAD. Fortunately, the Faraday effect caused by the magnetic field along the direction of light propagation can help to identify MAD from SANE when the polarization information is available. By comparing the rotation measure of jets from observations and simulations of MAD, \cite{2022ApJ...924..124Y} suggested that the accretion flow of M87 should be a MAD. Unfortunately, this method cannot be extensively applied to other low-luminosity active galactic nuclei (AGNs) because their polarization is usually weak (see \citealt{2022ApJ...924..124Y} for details). The more direct evidence for the presence of MAD was discovered recently in black hole X-ray binary MAXI J1820+070, where the radio fluxes are delayed about 8 days compared with the X-ray flux \citep{2023Sci...381..961Y}. In this process, the outburst of X-ray flux is produced by the expanding corona, resulting in the amplification of magnetic field and the formation of MAD around the time of the radio peak (8 days later).

Except for the MAD in an ADAF, it is also possible to generate a MAD in an optically thick and geometrically thin accretion disk (e.g., \citealt{2018MNRAS.480.3547M,2023arXiv230210226S}). By compiling a sample of radio-loud active galaxies, \citet{2014Natur.510..126Z} discovered a tight correlation between the jet magnetic flux and disk luminosity, which is consistent with the theoretical prediction of a MAD in MHD simulations. A similar relationship between the jet power and the disk luminosity supporting the MAD scenario is also reported in blazars \citep{2014Natur.515..376G}. However, the samples of these two works include both BL Lacertae objects and flat-spectrum radio quasars (FSRQs), which may correspond to either the MAD in an ADAF (BL Lac) or the MAD in a thin disk (FSRQs), respectively. Furthermore, their physical parameters adopted (e.g., jet power, magnetic flux and disk luminosity) are based on several approximations and assumptions. It is therefore difficult to separate MAD and SANE in a thin disk from these relationships.  

Another direct evidence for the MAD in a thin accretion disk comes from the obvious difference between the extreme ultraviolet (EUV) spectrum of radio-loud AGNs (RLAGNs) and the radio-quiet AGNs (RQAGNs, e.g., \citealt{2002ApJ...565..773T}), where the RLAGNs show a deficit in EUV band compared with the RQAGNs when the rest-frame wavelength is smaller than 1100 \AA. This phenomenon may indicate a missing or suppressed energy dissipation in the innermost region of accretion flow and  can be qualitatively explained by the MAD model in a thin disk \citep{2014ApJ...797L..33P,2015ApJ...806...47P}. For a thin disk with MAD in the innermost region, the presence of large-scale magnetic flux therein will extract a fraction of gas energy to the jet as Poynting flux, resulting in the observed EUV deficit. Furthermore, this picture can also qualitatively explain the positive relationship between the jet power normalized with the bolometric luminosity and the EUV deficit normalized to the RQAGN value shown by HST data (see \citealt{2015ApJ...806...47P} for details).

Except for the EUV spectrum, a significant difference in X-ray emission is also found between the MAD and SANE \citep{2022A&A...663L...4L}. The authors reported that the average X-ray flux of a MAD is 4.5 times higher than that of RQAGNs, while the average X-ray flux of RLAGNs without an EUV deficit is about 2–3 times larger than that of RQAGNs \citep{2013ApJ...763..109W,2018MNRAS.480.2861G}. The X-ray excess in MAD sample may originate from the increase of radiative efficiency in a corona, similar with the process that a MAD can improve the radiative efficiency of an ADAF (see \citealt{2022A&A...663L...4L} and the last section for details). This result implies that not all the RLAGNs contain a MAD. 
Unfortunately, such a statistically several times higher X-ray flux is not enough to discriminate between SANE and MAD given their error bars. Additionally, the method of utilizing the EUV deficit cannot be employed for most RLAGNs, as the data for EUV emission are often absent. Therefore, we turn our eyes to the radio band and find that all the objects in the MAD sample have strong radio emission (the radio-loudness are all higher than 100, see table 1 in \citealt{2022A&A...663L...4L}). However, it is unclear whether all the RLAGNs with strong radio emission contain a MAD. In this work, we intend to investigate this issue by comparing the multiwavelength properties of the 3CRR sources, the most powerful radio galaxies in the universe, and the MAD sample of \citet{2022A&A...663L...4L}.

\section{Sample}\label{sample}

\begin{center}
\begin{longtable}{lllllllllllll}
\caption{Sample.}\label{tab-data}\\
\hline
\multicolumn{1}{l}{Name} & \multicolumn{1}{l}{$z$} & \multicolumn{1}{l}{type} & \multicolumn{1}{l}{$\Gamma$} &  \multicolumn{1}{l}{log$L_{\rm {X}}$} & \multicolumn{1}{l}{Ref.} & \multicolumn{1}{l}{log$L_{\rm UV}$} &  \multicolumn{1}{l}{Ref.} & \multicolumn{1}{l}{$\Delta \log L_{\rm X}$} &
\multicolumn{1}{l}{log$L_{\rm R}$} & \multicolumn{1}{l}{$\log R_{\rm UV}$} & \multicolumn{1}{l}{log$\frac{M_{\rm bh}}{M_{\odot}}$} & \multicolumn{1}{l}{$\lambda$} \\

{} & {} & {} & {} & {[erg s$^{-1}$} & {} & {[erg s$^{-1}$} & {} & {[erg s$^{-1}$} & {[erg s$^{-1}$} & {} & {} & {}   \\
{} & {} & {} & {} & {Hz$^{-1}$]} & {} & {Hz$^{-1}$]} & {} & {Hz$^{-1}$]} & {Hz$^{-1}$]} & {} & {} & {}  \\
{(1)} &  {(2)} &  {(3)} &  {(4)} &
 {(5)} &  {(6)} & {(7)} &  {(8)} &  {(9)} &  {(10)} &  {(11)} &  {(12)} &  {(13)} \\
\hline
\endfirsthead

3C9   &  2.012  &  Q  &  1.61$^{+0.12}_{-0.08}$  &  27.56$^{+0.16}_{-0.12}$  &  1  &   31.46  & 8 &  0.14  &  32.92  &  1.46 & 9.8 & -0.70  \\
3C47   &  0.425  &  Q  &  1.87$^{+0.21}_{-0.22}$  &  27.31$^{+0.25}_{-0.17}$  &  1  &   30.00  & 8 &  1.01  &  32.60  &  2.61 & 9.2 & -1.56  \\
3C48   &  0.367  &  Q  &  2.32$^{+0.16}_{-0.01}$  &  27.16$^{+0.02}_{-0.02}$  &  1  &   30.81  & 8 &  0.24  &  33.55  &  2.74 & 9.2 & -0.75  \\
3C68.1   &  1.238  &  Q  &  1.9  &  27.21$^{+0.10}_{-0.10}$  &  1  &   30.43  & 8 &  0.57  &  31.81  &  1.38 & 9.9 & -1.83  \\
3C138   &  0.759  &  Q  &  1.46$^{+0.24}_{-0.11}$  &  27.57$^{+0.10}_{-0.10}$  &  1  &   30.29  & 8 &  1.04  &  33.27  &  2.98 & 8.9 & -0.97  \\
3C147   &  0.545  &  Q  &  1.85$^{+0.24}_{-0.22}$  &  26.91$^{+0.24}_{-0.15}$  &  1  &   30.33  & 8 &  0.35  &  34.38  &  4.05 & 9.1 & -1.13  \\
3C175   &  0.768  &  Q  &  1.52$^{+0.40}_{-0.07}$  &  27.52$^{+0.08}_{-0.07}$  &  1  &   31.03  & 8 &  0.42  &  32.68  &  1.65 & 9.9 & -1.22  \\
3C181   &  1.382  &  Q  &  1.72$^{+0.17}_{-0.10}$  &  27.56$^{+0.10}_{-0.08}$  &  1  &   30.84  & 8 &  0.61  &  32.66  &  1.82 & 9.6 & -1.12  \\
3C186   &  1.063  &  Q  &  1.88$^{+0.05}_{-0.03}$  &  27.29$^{+0.03}_{-0.03}$  &  1  &   30.89  & 8 &  0.30  &  32.80  &  1.91 & 9.5 & -0.96  \\
3C190   &  1.197  &  Q  &  1.61$^{+0.06}_{-0.05}$  &  27.22$^{+0.04}_{-0.04}$  &  1  &   30.51  & 8 &  0.53  &  33.60  &  3.09 & 8.7 & -0.55  \\
3C191   &  1.952  &  Q  &  1.68$^{+0.11}_{-0.10}$  &  27.64$^{+0.09}_{-0.07}$  &  1  &   30.75  & 8 &  0.76  &  33.82  &  3.07 & 9.7 & -1.31  \\
3C196   &  0.871  &  Q  &  1.67$^{+0.40}_{-0.38}$  &  27.23$^{+0.49}_{-0.25}$  &  1  &   30.48  & 8 &  0.56  &  32.28  &  1.80 & 9.6 & -1.48  \\
3C204   &  1.112  &  Q  &  2.15$^{+0.14}_{-0.22}$  &  27.75$^{+0.16}_{-0.14}$  &  1  &   31.07  & 8 &  0.63  &  33.10  &  2.03 & 9.5 & -0.79  \\
3C205   &  1.534  &  Q  &  1.83$^{+0.06}_{-0.17}$  &  27.93$^{+0.05}_{-0.07}$  &  1  &   31.38  & 8 &  0.57  &  33.28  &  1.90 & 9.6 & -0.58  \\
3C207   &  0.684  &  Q  &  1.59$^{+0.08}_{-0.23}$  &  27.39$^{+0.19}_{-0.15}$  &  1  &   30.22  & 8 &  0.91  &  33.91  &  3.69 & 8.5 & -0.64  \\
3C208   &  1.109  &  Q  &  1.63$^{+0.18}_{-0.14}$  &  27.44$^{+0.10}_{-0.08}$  &  1  &   30.99  & 8 &  0.38  &  33.37  &  2.39 & 9.4 & -0.77  \\
3C212   &  1.049  &  Q  &  1.85$^{+0.03}_{-0.08}$  &  27.85$^{+0.01}_{-0.01}$  &  1  &   30.02  & 8 &  1.52  &  33.79  &  3.77 & 9.2 & -1.54  \\
3C215  &  0.411  &  Q  &  1.88$^{+0.09}_{-0.08}$  &  27.16$^{+0.07}_{-0.05}$  &  1  &   29.65  & 8 &  1.12  &  31.92  &  2.27 & 8.3 & -1.01  \\
3C216  &  0.668  &  Q  &  1.72$^{+0.26}_{-0.20}$  &  27.15$^{+0.08}_{-0.09}$  &  1  &   29.85  & 8 &  0.96  &  34.20  &  4.35 & 8.6 & -1.11  \\
3C245  &  1.029  &  Q  &  1.57$^{+0.10}_{-0.05}$  &  27.63$^{+0.04}_{-0.03}$  &  1  &   30.19  & 8 &  1.17  &  34.55  &  4.37 & 9.4 & -1.57  \\
3C249.1  &  0.311  &  Q  &  1.92$^{+0.13}_{-0.04}$  &  26.85$^{+0.08}_{-0.08}$  &  1  &   30.25  & 8 &  0.35  &  32.29  &  2.04 & 9.3 & -1.40  \\
3C254  &  0.734  &  Q  &  1.93$^{+0.10}_{-0.07}$  &  27.36$^{+0.02}_{-0.02}$  &  1  &   30.70  & 8 &  0.52  &  32.55  &  1.84 & 9.3 & -0.95  \\
3C263  &  0.652  &  Q  &  1.83$^{+0.09}_{-0.05}$  &  27.27$^{+0.01}_{-0.01}$  &  1  &   30.88  & 8 &  0.30  &  33.35  &  2.47 & 9.1 & -0.58  \\
3C268.4  &  1.400  &  Q  &  1.45$^{+0.17}_{-0.12}$  &  27.62$^{+0.11}_{-0.09}$  &  1  &   31.20  & 8 &  0.40  &  33.59  &  2.39 & 9.8 & -0.96  \\
3C270.1  &  1.519  &  Q  &  1.57$^{+0.07}_{-0.07}$  &  27.29$^{+0.03}_{-0.03}$  &  1  &   30.87  & 8 &  0.83  &  34.25  &  3.38 & 9.0 & -0.49  \\
3C275.1  &  0.557  &  Q  &  1.79$^{+0.11}_{-0.10}$  &  27.02$^{+0.06}_{-0.06}$  &  1  &   29.77  & 8 &  0.89  &  33.11  &  3.35 & 8.3 & -0.89  \\
3C286  &  0.849  &  Q  &  2.12$^{+0.51}_{-0.26}$  &  27.09$^{+0.12}_{-0.10}$  &  1  &   30.75  & 8 &  0.22  &  35.15  &  4.41 & 8.5 & -0.11  \\
3C287  &  1.055  &  Q  &  1.86$^{+0.04}_{-0.04}$  &  27.39$^{+0.03}_{-0.03}$  &  1  &   30.59  & 8 &  0.64  &  35.10  &  4.51 & 9.6 & -1.37  \\
3C309.1  &  0.904  &  Q  &  1.57$^{+0.05}_{-0.03}$  &  27.74$^{+0.05}_{-0.03}$  &  1  &   30.81  & 8 &  0.82  &  34.84  &  4.03 & 9.1 & -0.65  \\
3C325  &  0.860  &  Q  &  1.45$^{+0.07}_{-0.02}$  &  26.54$^{+0.02}_{-0.01}$  &  1  &   29.84  & 8 &  0.36  &  31.80  &  1.96 & 9.6 & -2.12  \\
3C334  &  0.555  &  Q  &  1.85$^{+0.18}_{-0.11}$  &  27.12$^{+0.04}_{-0.04}$  &  1  &   30.56  & 8 &  0.39  &  33.04  &  2.48 & 9.7 & -1.50  \\
3C336  &  0.927  &  Q  &  1.93$^{+0.24}_{-0.28}$  &  27.49$^{+0.14}_{-0.11}$  &  1  &   30.58  & 8 &  0.74  &  32.80  &  2.22  & 9.2 & -0.98 \\
3C345  &  0.594  &  Q  &  1.69$^{+0.09}_{-0.02}$  &  27.66$^{+0.02}_{-0.02}$  &  1  &   30.22  & 8 &  1.18  &  35.00  &  4.78 & 9.3 & -1.44  \\
3C351  &  0.371  &  Q  &  1.90$^{+0.20}_{-0.13}$  &  27.02$^{+0.17}_{-0.09}$  &  1  &   30.52  & 8 &  0.32 &  31.42  &  0.90 & 9.5 & -1.33  \\
4C16.49  &  1.296  &  Q  &  1.83$^{+0.20}_{-0.12}$  &  27.49$^{+0.10}_{-0.09}$  &  1  &   30.59  & 8 &  0.74 &  33.02  &  2.43 & 9.8 & -1.57  \\
3C20 &  0.174  &  R  &  1.65$^{+ 0.21}_{-0.19}$  &  25.99${\pm 0.10}$  &  2  &   27.24  &  5  &  1.78  &  30.27  &  3.03 & 8.7 & -2.44  \\
3C33.1 &  0.181  &  R  &  0.89$^{+ 0.33}_{-0.37}$  &  26.01${\pm 0.13}$  &  2  &   28.37  &  6  &  0.94  &  30.98  &  2.61 & 8.6 & -1.58  \\
3C61.1   &  0.188  &  R  &  1.70  &  25.90${\pm 0.13}$  &  2  &   26.81  &  5  &  2.02   &  30.31  &  3.50 & 8.1 & -0.91   \\
3C79   &  0.256  &  R  &  1.77$^{+ 0.69}_{-0.61}$  &  25.78${\pm 0.58}$  &  3  &   28.87  &  7  &  0.33  &  30.82  &  1.96 & 9.0 & -1.42   \\
3C171  &  0.24  &  R  &  1.67$^{+ 0.20}_{-0.23}$  &  25.67 
${\pm 0.11}$  &  2  &   26.79  &  5  &  1.80  &  30.13  &  3.34 & 8.7 & -1.09    \\
3C184.1&  0.118  &  R  &  0.57$^{+ 0.30}_{-0.34}$  &  25.39${\pm 0.16}$  &  2  &   28.50  &  6  &  0.23  &  30.28  &  1.78 &  8.5 & -1.55  \\
3C192  &  0.0597  &  R  &  1.70  &  24.89 ${\pm 0.19}$  &  3  &   27.33  &  5   &  0.61   &  29.80  &  2.47 & 8.7 & -2.64   \\
3C219  &  0.175  &  R  &  1.90$^{+ 0.19}_{-0.18}$  &  26.03${\pm 0.07}$  &  3  &   28.51  &  6  &  0.86  &  31.57  &  3.06 &  9.2 & -2.71 \\
3C223  &  0.137  &  R  &  0.71$^{+ 0.77}_{-0.61}$  &  25.21${\pm 0.60}$  &  3  &   27.67  &  6  &  0.67  &  30.59  &  2.92 &  8.7 & -1.81   \\
3C234  &  0.185  &  R  &  1.39$^{+ 0.10}_{-0.16}$  &  26.21${\pm 0.09}$  &  2  &   29.48  &  6  &  0.30  &  31.87  &  2.40 &   9.3 & -1.47  \\
3C285  &  0.0794 &  R  &  1.70  &  25.30${\pm 0.10}$  &  3  &   26.82  &  6  &  1.41  &  29.93  &  3.11 & 8.6 & -3.33   \\
3C300  &  0.270  &  R  &  1.78${\pm 0.06}$  &  25.39${\pm 0.02}$  &  2  &   28.04  &  6  &  0.57  &  31.23  &  3.19 & 8.8 & -2.07   \\
3C303  &  0.141  &  R  &  1.60${\pm 0.21}$  &  25.84${\pm 0.08}$  &  3  &   28.77  &  6  &  0.46  &  31.84  &  3.07 & 9.0 & -2.54   \\
3C321  &  0.0965 &  R  &  1.70  &  24.06${\pm 1.32}$  &  4  &   27.28  &  6  &  -0.18 &  30.80  &  3.52 & 9.1 & -3.47   \\
\hline
\multicolumn{13}{p{18cm}} {Notes: Col. (1): Source name. Col. (2): Redshift. Col. (3): AGN type. Q: quasar; R: radio galaxy. Col. (4): X-ray photon index.
Col.(5): X-ray intrinsic luminosity at 2 keV. Col. (6): References for $L_{\rm {X}}$. Col. (7): optical-UV luminosity at 2500 \AA. 
Col. (8): References  for $L_{\rm {UV}}$. Col. (9): Excess of X-ray luminosity in 3CRR sources to that in RQAGNs. 
Col. (10): Radio spectral luminosity at 5 GHz. Col. (11): Radio loudness defined as $R_{\rm {UV}}=L_{\rm {R}}/L_{\rm {UV}}$.
Col. (12): black hole mass. Col. (13): Eddington ratio. }\\
\multicolumn{13}{p{18cm}}{References. 1, \citet{2020ApJ...893...39Zhou}; 2, \citet{2009MNRAS.396.1929H}; 3, \citet{2006MNRAS.370.1893H}; 4, \citet{2006ApJ...642...96E}; 5, \citet{2006ApJS..164..307M}; 6, \citet{2010ApJ...725.2426B}; 7, \citet{2008ApJ...677...79F}; 8, this work. }
\end{longtable}
\end{center}



The RLAGNs with EUV deficit are suggested to contain a thin accretion disk with MAD in the inner region \citep{2014ApJ...797L..33P,2015ApJ...806...47P}. In order to compare consistently with the MAD sample, the accretion mode of 3CRR sources is also required to be a thin disk. Therefore, we adopt the quasars and the high-excitation radio galaxies (HERGs) in 3CRR catalog in this work. For one thing, the accretion flow of quasar is widely accepted to be a radiatively efficient thin disk due to the big-blue-bump in optical-UV band, which is thought to originate from an optically thick, geometrically thin disk. For another, \citet{2010A&A...509A...6B} suggested that a radio galaxy with excitation index ($\rm EI = \log[O_{\rm III}]/H_{\rm \beta}-1/3\left(log[N_{\rm II}]/H_{\rm \alpha}+ log[S_{\rm II}]/H_{\rm \alpha} + log[O_{\rm I}]/H_{\rm \alpha}\right)$) higher than 0.95 is a HERG, whose Eddington ratio is found to be higher than 0.01 and is suggested to comprise a thin accretion disk around the black hole (e.g., \citealt{2012MNRAS.421.1569B}). 

3CRR catalog contains 43 quasars and 130 radio galaxies. Firstly, we pick up all the quasars and HERGs with detected EI larger than 0.95 in 3CRR catalog, where the values of EI are given by \citet{2010A&A...509A...6B} and \citet{2016RAA....16..136H}. A sample of 43 quasars and 17 radio galaxies are obtained. To avoid uncertainty from different telescopes, only sources observed by Hubble Space Telescope (HST) are chosen, since the optical-UV emission of MAD sample is detected by HST \citep{2015ApJ...806...47P,2022A&A...663L...4L}. The optical-UV data of 3CRR catalog observed with HST has been extensively explored in many previous works (e.g.,\citealt{2010A&A...509A...6B,2016AJ....151..120W,2023ApJ...945..145A,2023NatAs...7..262K}). 

In order to obtain their properties in a consistent way, we analyze the HST images of all the 43 quasars ourselves using the Python package photulis \citep{Bradley2020}. First, we estimate the background level and its standard deviation $\sigma$, and sigma clip the background data with a threshold of 2$\sigma$. 
Following this, we use the task detect$\_$sources to detect sources and generate a segmentation map, providing information of sources including their centroid, semi-major and semi-minor axis, as well as position angle (PA), etc. Using this information, we can derive the surface brightness profile of the target with the task ellipse. By maintaining the ellipticity and PA obtained initially, we refit the isophotes in linear steps to sample the target's surface brightness profile. Finally, we determine the aperture size by the isophote at the surface brightness level of 25 $\rm mag\ arcsec^{-2}$, and perform the force aperture photometry to estimate the flux and uncertainty of the target. By visually checking the fitting results and analyzing the ratio of the obtained flux to the corresponding uncertainty (SNR), we have adopted the results of 35 quasars with SNR larger than 10. The flux at different optical-UV wavelengths (eg. 6060 \AA \space of F606W filter, 4750 \AA \space of F475W filter) are obtained from photometry analysis of the HST images, and then de-reddened for Galactic extinction using the \cite{Cardelli_etal_1989} Milky Way reddening law and E(B-V) derived from the \cite{Schlegel_etal_1998} dust map. The fluxes at 2500 \AA\ are then calculated by extrapolating the HST optical-UV flux and assuming a spectral index $\alpha_o = -0.5$ ($f_\nu\sim\nu^{\alpha_o}$, e.g.,  \citealt{2011ApJS..196....2S}). We fit only the quasars because it is difficult to detect the weak emission lines needed to calculate EI for middle and high redshift HERGs. Additionally, it is challenging to achieve a high SNR in middle and high redshift (z > 0.5) HERG observations due to their faint flux compared to background noise.


Lastly, we get a 3CRR sample composed of 35 quasars and 14 HERGs (see table 1), where all objects have been observed with Chandra and/or XMM-Newton telescope, so the 2 keV X-ray flux is avaible. The photon index $\Gamma$ in Col. (4) and the intrinsic X-ray luminosity at 2 keV of quasars ($L_{\rm X}$, obtained from X-ray spectral
modeling) in Col. (5) are all taken from \citet{2020ApJ...893...39Zhou}, while the $L_{\rm X}$ of HERGs are from NED\footnote{ http://ned.ipac.caltech.edu/} (see Col. 6 for references). All the $L_{\rm X}$ have been corrected for both intrinsic and Milky Way $N_{\rm H}$ and K-corrected. We can find that, except for three radio galaxies (3C79, 3C223 and 3C321), the uncertainty of X-ray luminosity in other sources are all quite small. 
The spectral luminosity at 2500 \AA ($L_{\rm UV}$) of HERGs in Col. (7) are also from NED (see Col. (8) for detailed references), which is derived from the data at nearby frequency assuming a spectral index $\alpha_{\rm o}=-0.5$ (\citealt{2011ApJS..196....2S}). The $\Delta {\rm log} L_{\rm X}$ in Col. (9) indicates the excess of X-ray luminosity above the RQAGN value, calculated from the $L_{\rm UV}-L_{\rm X}$ relationship of \citet{2010A&A...512A..34L} (for further details see Section \ref{results}). The core radio luminosity at 5 GHz in Col. (10) is taken from the online 3CRR catalog\footnote{https://3crr.extragalactic.info/}. Except for 3C216, 3C325, and 3C345, the black hole mass of quasars and radio galaxies in Col. (12) are all taken from \citet{2006MNRAS.368.1395M} and \citet{2010A&A...509A...6B}, respectively. The black hole mass in 3C216, 3C325, and 3C345 are obtained from  \citet{2016AJ....151..120W}, \citet{2023ApJ...945..145A}, and \citet{2011ApJS..194...45S}, respectively. In quasars, the broad line width of H$\beta$, \ion{C}{4}, and \ion{Mg}{2} is separately adopted to calculate the black hole mass (see \citealt{2006MNRAS.368.1395M} for details), while the host galaxy bulge H band luminosity is used in radio galaxies (see \citealt{2010A&A...509A...6B} for details). These two methods give consistent BH mass estimates (see Fig. 1 in \citealt{2021A&A...654A.141L}). The Eddington ratio of quasars and radio galaxies in Col. (13) are also estimated by different methods. We use the bolometric correction factor provided by \citet{2012MNRAS.422..478R} to calculate the bolometric luminosity of quasars, while the calculation of radio galaxy bolometric luminosity is based on the [OIII]$\lambda$5007 luminosity  \citep{2010A&A...509A...6B}, as the correction factor for radio galaxies is not available. Indeed, [OIII] luminosity may not be as a reliable proxy for bolometric luminosity for the HERGs as it is for the quasars. This is because the [OIII] emitting region extends to sufficiently small radius (i.e., sufficiently close to the supermassive black hole) to be partly obscured by the dusty torus when the active nucleus is viewed edge-on (see, e.g., \citealt{1990Natur.343...43J,1997MNRAS.286...23B,2006A&A...453..525N,2014ApJ...788...98D}).

\section{Results}\label{results}
\begin{figure}
\centering
\includegraphics[width=15cm]{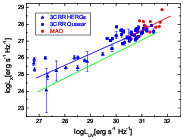}
\caption{Relationship between the X-ray luminosity $L_{\rm X}$ at 2 keV and optical-UV luminosity $L_{\rm UV}$ at 2500 \AA\ in 3CRR sample, where the red circles, blue squares and blue triangles correspond to the MAD, 3CRR quasar and 3CRR HERGs, respectively. The blue line represents our best-fitting result for 3CRR sources. The green line and the red line are the best-fitting result for RQAGNs given by \citet{2010A&A...512A..34L} and the relationship for MAD sources given by \citet{2022A&A...663L...4L}, respectively. }\label{f1}
\end{figure}

\begin{figure}
\centering
\includegraphics[width=15cm]{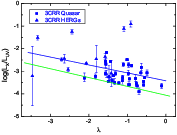}
\caption{Relationship between $L_{\rm X}/L_{\rm UV}$ and Eddington ratio $\lambda$ in 3CRR sample, where the blue and green lines are the best-fitting results for the 3CRR sample and the RQAGN sample given by \citet{2010A&A...512A..34L}, respectively. }\label{f12}
\end{figure}

\begin{figure}
\centering
\includegraphics[width=15cm]{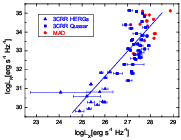}
\caption{Relationship between the radio luminosity $L_{\rm R}$ at 5 GHz and the X-ray luminosity $L_{\rm X}$ at 2 keV, where the blue line represents the best-fitting result to the 3CRR sources. }\label{f2}
\end{figure}

\begin{figure}
\centering
\includegraphics[width=15cm]{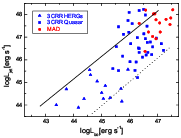}
\caption{Relationship between the jet power $L_{\rm jet}$ and the bolometric luminosity $L_{\rm bol}$, where the solid line and the dotted line represent the maximum jet efficiency ($\eta=140\%$ with $a=0.99$) and the jet efficiency $\eta=1.4\%$ with $a=0.1$, respectively \citep{2011MNRAS.418L..79T}.} \label{f3}
\end{figure}

$L_{\rm X}-L_{\rm UV}$ relationship is extensively adopted to investigate the accretion physics in quasars. \citet{2010A&A...512A..34L} presented a tight relationship between $L_{\rm X}$ and $L_{\rm UV}$, i.e., $\log L_{\rm X}=(0.76\pm 0.02) \log L_{\rm UV} +3.51\pm 0.64$ by compiling a type-1 RQAGN sample. This relationship can be qualitatively fitted with the disk-corona model \citep{2017A&A...602A..79L,2018MNRAS.477..210Q}, where the X-ray emission comes from the inverse Compton scattering
of optical-UV photons from a thin disk by the hot corona. However, though the average X-ray luminosity of MAD sample is found to be roughly 4.5 times larger than that in RQAGNs, matched in $L_{\rm UV}$, their slopes are very similar (the slope is 0.75, see \citealt{2022A&A...663L...4L} for details), which may also indicate a disk-corona origin of the $L_{\rm X}-L_{\rm UV}$ relationship in MAD objects. 

Not all the RLAGNs contain a MAD since the average X-ray luminosity is only 2-3 times higher than that in RQAGNs \citep{2018MNRAS.480.2861G,2021A&A...654A.141L}. Therefore, in order to check whether all the 3CRR sources with a thin disk and strong radio emissions contain a MAD, we firstly compare their  $L_{\rm X}-L_{\rm UV}$ relationship with the MAD sample given by \citet{2022A&A...663L...4L} in Fig. \ref{f1}, where the blue line represents our best-fitting result. The red and green lines are the best-fitting results for MAD sample \citep{2022A&A...663L...4L} and RQAGNs \citep{2010A&A...512A..34L}, respectively. It is found that the average X-ray flux of 3CRR sources is about 5.0 times higher than that of RQAGNs, which is consistent with the 4.5 times given in MAD sample within uncertainty. We adopt the ordinary least squares (OLS) method to fit all the relationships for both the 3CRR and the MAD samples \citep{1990ApJ...364..104I}. The correlation between $L_{\rm X}$ and $L_{\rm UV}$ reads
\begin{equation}
    \log L_{\rm X}=(0.68\pm 0.06) \log L_{\rm UV} +6.51\pm 2.02,
\end{equation}
where the confidence level based on a Pearson test is larger than 99.9\%. The slope $0.68\pm 0.06$ of $L_{\rm X}-L_{\rm UV}$ relationship is shallower but consistent within errors with that in MAD sample ($0.75\pm 0.25$). However, it is known that the ratio of X-ray luminosity to the UV luminosity will increase with the decreasing Eddington ratio (see e.g., \citealt{2010A&A...512A..34L}). Therefore, we check the distribution of Eddington ratio in our sample and find it is equally distributed (Fig. \ref{f12}). Our best-fitting result (the blue line) is several times higher than that from RQAGNs in \citet{2010A&A...512A..34L} (the green line), and the confidence level of the relationship
is 99.3\%. We note that many HERGs have  $\lambda < 0.01 $ in our sample, which marks the division between a thin disk and a radiatively inefficient accretion flow \citep{2014ARA&A..52..529Y}. However, similar values for HERGs had been reported by other works with larger samples (see e.g., Fig. 11 in \citealt{2010A&A...509A...6B} and Fig. 6 in \citealt{2012MNRAS.421.1569B}). The reasons for this may stem from the significant uncertainties associated with calculating both the black hole mass using bulge H-band luminosity and the bolometric luminosity using [OIII] luminosity.

The traditional relationship between radio (5GHz, $L_{\rm R}$) and X-ray (2 keV, $L_{\rm X}$) luminosities in AGNs is also investigated in Fig. \ref{f2}, where the blue line is the best-fitting result, which can be given as 
\begin{equation}
\log L_{\rm R}=(1.35\pm 0.11) \log L_{\rm X} -3.47\pm 3.11, 
\end{equation}
where the confidence level based on a Pearson test is larger than 99.9\%. Though the MAD sample is located on the upper right corner of 3CRR objects due to the higher luminosity, they follow the $L_{\rm R}-L_{\rm X}$ relationship (the blue line) in 3CRR objects quite well. The results of Fig. \ref{f1} and \ref{f2} indicate that all the quasars and many HERGs in the 3CRR catalog may contain a MAD. 


In addition, the GRMHD simulations suggested that the jet efficiency ($\eta= L_{\rm BZ}/\dot{M}c^2$) in a MAD could be very large, where $L_{\rm BZ}$ and $\dot{M}$ are the jet power produce by the BZ mechanism \citep{1977MNRAS.179..433B} and the mass accretion rate, respectively. The jet efficiency $\eta$ can be larger than 1 in MAD around a black hole with high spin (e.g., \citealt{2011MNRAS.418L..79T,2012MNRAS.423.3083M}), where the extra energy is suggested to be extracted from the rotating black hole. In Fig. \ref{f3}, we investigate the relationship between the jet power $L_{\rm jet}$ and the bolometric luminosity $L_{\rm bol}$ for both the 3CRR sample and MAD sample. Here $L_{\rm jet}$ ($=0.81 \log L_{\rm R} + 11.9$) is the kinetic jet power calculated with the radio core luminosity at 5 GHz \citep{2007MNRAS.381..589M}. The black solid line represents the maximum jet efficiency $\eta=140\%$ in a MAD with black hole spin $a=0.99$ and the dotted line is for $\eta=1.4\%$, which roughly corresponds to $a=0.1$ since $L_{\rm BZ}\propto a^2$ \citep{2011MNRAS.418L..79T}. It is found that most of the 3CRR sources and MAD sample lie between the solid and dotted lines. Therefore, the jet efficiencies of 3CRR sources are consistent with the GRMHD simulations of MAD.

\section{Discussion} \label{discussion}

In theory, the formation of MAD depends on the accumulation of large-scale magnetic fields in the inner disk region. For a standard thin disk, however, it is difficult to amplify the large-scale magnetic field due to its low radial velocity, resulting in the relatively faster diffusion of magnetic field compared with advection (e.g., \citealt{1994MNRAS.267..235L}). However, this problem may be resolved by taking the magnetically driven outflow into account, which can help to increase the radial velocity of gas by transferring the angular momentum of disk \citep{2013ApJ...765..149C,2014ApJ...786....6L}.  Furthermore, the outflow takes away not only the angular momentum, but also the energy of gas, resulting in a far smaller radiative efficiency. The radiative efficiency given by theoretical model can be several order of magnitudes smaller than 10\% in extreme cases (see e.g.,  \citealt{2014ApJ...788...71L, 2019ApJ...886...92L, 2023MNRAS.526..862Z}). This result can explain the phenomenon that the jet efficiency of some objects in Fig. \ref{f3} are even several times higher than 140\%. The detailed physics of a disk-corona system with MAD is still unclear, though the MHD simulations do suggest a higher jet power and smaller bolometric luminosity \citep{2011MNRAS.418L..79T,2012MNRAS.423.3083M,2018MNRAS.480.3547M}. However, for an ADAF, the presence of MAD can increase its radiative efficiency \citep{2019ApJ...887..167X}. If the properties of corona is similar with ADAF as suggested by some works \citep{2000A&A...361..175M,2007ApJ...671..695L,2018MNRAS.477..210Q}, we can anticipate that the X-ray flux will also increase several times for a MAD in thin disk.

Recent works suggest that the magnetically elevated accretion disks (e.g., \citealt{2022MNRAS.511.2040B, 2023MNRAS.521.5952B}) can also lead to a MAD. In this scenario, the authors suggested that it is the strong toroidal magnetic field making an accretion disk MAD, though weak poloidal field is still necessary to trigger the dynamo process. However, the jet power in a magnetically elevated accretion disk should be much weaker compared with the standard MAD scenario, since the poloidal field responsible for the jet power is weak ($L_{\rm BZ}\propto B_{\rm p}^2$, see e.g., \citealt{1997MNRAS.292..887G}). 

The 3CRR catalog contains objects with the strongest radio emissions at 178 MHz in the sky, so the black hole spins in these objects are possibly high, leading to high jet efficiencies. Although the jet in the objects around the dotted line in Fig. \ref{f3} can be launched by a SANE with high black hole spin, other objects around the solid line should contain a MAD judging by their high jet efficiency. Furthermore, the EUV deficit in the MAD sample is difficult to explain with the SANE scenario. The origin of the X-ray luminosity excess remains debatable. In principle, both the jet core and the corona, influenced by the large-scale magnetic field, can contribute to this excess. However, if this excess were attributed to the jet core, two specific conditions must be met to maintain a similar $L_{\rm X}$-$L_{\rm UV}$ slope between MAD and RQAGNs: 1) the jet core must depend strongly on $L_{\rm UV}$; and 2) the slope of this dependence must be consistent with the $L_{\rm X}$-$L_{\rm UV}$ slope in RQAGNs (see \citealt{2020MNRAS.496..245Zhu} for details). Therefore, it is more natural to assume a coronal origin for the X-ray emissions. However, the jet core and corona may be indistinguishable in spatial resolution. In this context, one could also argue that the X-ray luminosity excess originates from the jet. This slope is also consistent with that observed in radio-loud quasars (see e.g., \citealt{2020MNRAS.496..245Zhu,2021A&A...654A.141L}).

\section{Summary} \label{summary}

We investigate whether all RLAGNs with strong radio emission harbor a MAD by compiling a sample of quasars and HERGs from 3CRR catalog in this work. By comparing the multi-band data with the MAD sample given in \citet{2022A&A...663L...4L}, it is found that the average X-ray luminosity of 3CRR objects is higher by the same factor as the MAD objects when compared with the RQAGNs matched in $L_{\rm UV}$ from \citet{2010A&A...512A..34L} (Fig. \ref{f1}). The  $L_{\rm X}-L_{\rm UV}$ slope of the 3CRR sample is also consistent with that of the MAD sample. Furthermore, the 3CRR sample has a larger range of $L_{\rm X}$ (4 dex) than the MAD sample (1 dex), so the $L_{\rm X}-L_{\rm UV}$ relation and slope are much better determined than those from the MAD sample. We find a positive relationship between the radio (5 GHz) luminosity $L_{\rm R}$ and the 2 keV X-ray luminosity $L_{\rm X}$ for the 3CRR sample (Fig. \ref{f2}). The MAD sample is found to follow the $L_{\rm R}-L_{\rm X}$ relationship quite well, which further support the same physical mechanism for MAD sample and 3CRR sources. Lastly, the jet efficiencies of the 3CRR sources (Fig. \ref{f3}) are also found to be qualitatively consistent with GRMHD simulations of MAD. These results indicate that all the quasars and a fraction of HERGs (at least for z<0.3) in 3CRR catalog should contain a MAD.

\section* {Acknowledgements}
We thank the reviewer for very valuable and detailed comments, which have greatly improved the quality of this manuscript. This work is supported by the NSFC (grants 12273089, 12073023, 12233007, 12361131579, 12347103), the science research grants from the China Manned Space Project with No. CMSCSST-2021-A06, and the Fundamental Research Fund for Chinese Central Universities. This work has made extensive use of the NASA/IPAC Extragalactic Database (NED) and the 3CRR catalog \citep{1983MNRAS.204..151L}. NED is operated by the Jet Propulsion Laboratory, California Institute of Technology, under contract with the National Aeronautics and Space Administration (NASA). This research has also made extensive use of data obtained from the Chandra Data Archive and the observations made with the NASA/ESA Hubble Space Telescope, obtained from the Hubble Legacy Archive, which is a collaboration between the Space Telescope Science Institute (STScI/NASA), the Space Telescope European Coordinating Facility (ST-ECF/ESAC/ESA) and the Canadian Astronomy Data Centre (CADC/NRC/CSA).

\bibliography{3CRR}

\begin{thebibliography}{}
\expandafter\ifx\csname natexlab\endcsname\relax\def\natexlab#1{#1}\fi
\providecommand{\url}[1]{\href{#1}{#1}}
\providecommand{\dodoi}[1]{doi:~\href{http://doi.org/#1}{\nolinkurl{#1}}}
\providecommand{\doeprint}[1]{\href{http://ascl.net/#1}{\nolinkurl{http://ascl.net/#1}}}
\providecommand{\doarXiv}[1]{\href{https://arxiv.org/abs/#1}{\nolinkurl{https://arxiv.org/abs/#1}}}

\bibitem[{{Azadi} {et~al.}(2023){Azadi}, {Wilkes}, {Kuraszkiewicz}, {McDowell},
  {Siebenmorgen}, {Ashby}, {Birkinshaw}, {Worrall}, {Abrams}, {Barthel},
  {Fazio}, {Haas}, {Hyman}, {Mart{\'\i}nez-Galarza}, \&
  {Meyer}}]{2023ApJ...945..145A}
{Azadi}, M., {Wilkes}, B., {Kuraszkiewicz}, J., {et~al.} 2023, \apj, 945, 145,
  \dodoi{10.3847/1538-4357/acbe9c}

\bibitem[{{Baker}(1997)}]{1997MNRAS.286...23B}
{Baker}, J.~C. 1997, \mnras, 286, 23, \dodoi{10.1093/mnras/286.1.23}

\bibitem[{{Baldi} {et~al.}(2010){Baldi}, {Chiaberge}, {Capetti}, {Sparks},
  {Macchetto}, {O'Dea}, {Axon}, {Baum}, \& {Quillen}}]{2010ApJ...725.2426B}
{Baldi}, R.~D., {Chiaberge}, M., {Capetti}, A., {et~al.} 2010, \apj, 725, 2426,
  \dodoi{10.1088/0004-637X/725/2/2426}

\bibitem[{{Begelman} \& {Armitage}(2023)}]{2023MNRAS.521.5952B}
{Begelman}, M.~C., \& {Armitage}, P.~J. 2023, \mnras, 521, 5952,
  \dodoi{10.1093/mnras/stad914}

\bibitem[{{Begelman} {et~al.}(2022){Begelman}, {Scepi}, \&
  {Dexter}}]{2022MNRAS.511.2040B}
{Begelman}, M.~C., {Scepi}, N., \& {Dexter}, J. 2022, \mnras, 511, 2040,
  \dodoi{10.1093/mnras/stab3790}

\bibitem[{{Best} \& {Heckman}(2012)}]{2012MNRAS.421.1569B}
{Best}, P.~N., \& {Heckman}, T.~M. 2012, \mnras, 421, 1569,
  \dodoi{10.1111/j.1365-2966.2012.20414.x}

\bibitem[{{Blandford} \& {Znajek}(1977)}]{1977MNRAS.179..433B}
{Blandford}, R.~D., \& {Znajek}, R.~L. 1977, \mnras, 179, 433,
  \dodoi{10.1093/mnras/179.3.433}

\bibitem[{{Bradley} {et~al.}(2020){Bradley}, {Sip{\H{o}}cz}, {Robitaille},
  {Tollerud}, {Vin{\'\i}cius}, {Deil}, {Barbary}, {Wilson}, {Busko},
  {G{\"u}nther}, {Cara}, {Conseil}, {Bostroem}, {Droettboom}, {Bray}, {Andersen
  Bratholm}, {Lim}, {Barentsen}, {Craig}, {Pascual}, {Perren}, {Greco},
  {Donath}, {De Val-Borro}, {Kerzendorf}, {Bach}, {Weaver}, {D'Eugenio},
  {Souchereau}, \& {Ferreira}}]{Bradley2020}
{Bradley}, L., {Sip{\H{o}}cz}, B., {Robitaille}, T., {et~al.} 2020,
  {astropy/photutils: 1.0.1}, 1.0.1,  Zenodo, \dodoi{10.5281/zenodo.4049061}

\bibitem[{{Buttiglione} {et~al.}(2010){Buttiglione}, {Capetti}, {Celotti},
  {Axon}, {Chiaberge}, {Macchetto}, \& {Sparks}}]{2010A&A...509A...6B}
{Buttiglione}, S., {Capetti}, A., {Celotti}, A., {et~al.} 2010, \aap, 509, A6,
  \dodoi{10.1051/0004-6361/200913290}

\bibitem[{{Cao} \& {Spruit}(2013)}]{2013ApJ...765..149C}
{Cao}, X., \& {Spruit}, H.~C. 2013, \apj, 765, 149,
  \dodoi{10.1088/0004-637X/765/2/149}

\bibitem[{{Cardelli} {et~al.}(1989){Cardelli}, {Clayton}, \&
  {Mathis}}]{Cardelli_etal_1989}
{Cardelli}, J.~A., {Clayton}, G.~C., \& {Mathis}, J.~S. 1989, \apj, 345, 245,
  \dodoi{10.1086/167900}

\bibitem[{{Chatterjee} \& {Narayan}(2022)}]{2022ApJ...941...30C}
{Chatterjee}, K., \& {Narayan}, R. 2022, \apj, 941, 30,
  \dodoi{10.3847/1538-4357/ac9d97}

\bibitem[{{Dicken} {et~al.}(2014){Dicken}, {Tadhunter}, {Morganti}, {Axon},
  {Robinson}, {Magagnoli}, {Kharb}, {Ramos Almeida}, {Mingo}, {Hardcastle},
  {Nesvadba}, {Singh}, {Kouwenhoven}, {Rose}, {Spoon}, {Inskip}, \&
  {Holt}}]{2014ApJ...788...98D}
{Dicken}, D., {Tadhunter}, C., {Morganti}, R., {et~al.} 2014, \apj, 788, 98,
  \dodoi{10.1088/0004-637X/788/2/98}

\bibitem[{{Evans} {et~al.}(2006){Evans}, {Worrall}, {Hardcastle}, {Kraft}, \&
  {Birkinshaw}}]{2006ApJ...642...96E}
{Evans}, D.~A., {Worrall}, D.~M., {Hardcastle}, M.~J., {Kraft}, R.~P., \&
  {Birkinshaw}, M. 2006, \apj, 642, 96, \dodoi{10.1086/500658}

\bibitem[{{Fu} \& {Stockton}(2008)}]{2008ApJ...677...79F}
{Fu}, H., \& {Stockton}, A. 2008, \apj, 677, 79, \dodoi{10.1086/529015}

\bibitem[{{Ghisellini} {et~al.}(2014){Ghisellini}, {Tavecchio}, {Maraschi},
  {Celotti}, \& {Sbarrato}}]{2014Natur.515..376G}
{Ghisellini}, G., {Tavecchio}, F., {Maraschi}, L., {Celotti}, A., \&
  {Sbarrato}, T. 2014, \nat, 515, 376, \dodoi{10.1038/nature13856}

\bibitem[{{Ghosh} \& {Abramowicz}(1997)}]{1997MNRAS.292..887G}
{Ghosh}, P., \& {Abramowicz}, M.~A. 1997, \mnras, 292, 887,
  \dodoi{10.1093/mnras/292.4.887}

\bibitem[{{Gupta} {et~al.}(2018){Gupta}, {Sikora}, {Rusinek}, \&
  {Madejski}}]{2018MNRAS.480.2861G}
{Gupta}, M., {Sikora}, M., {Rusinek}, K., \& {Madejski}, G.~M. 2018, \mnras,
  480, 2861, \dodoi{10.1093/mnras/sty2043}

\bibitem[{{Hardcastle} {et~al.}(2006){Hardcastle}, {Evans}, \&
  {Croston}}]{2006MNRAS.370.1893H}
{Hardcastle}, M.~J., {Evans}, D.~A., \& {Croston}, J.~H. 2006, \mnras, 370,
  1893, \dodoi{10.1111/j.1365-2966.2006.10615.x}

\bibitem[{{Hardcastle} {et~al.}(2009){Hardcastle}, {Evans}, \&
  {Croston}}]{2009MNRAS.396.1929H}
---. 2009, \mnras, 396, 1929, \dodoi{10.1111/j.1365-2966.2009.14887.x}

\bibitem[{{Hu} {et~al.}(2016){Hu}, {Cao}, {Chen}, \&
  {You}}]{2016RAA....16..136H}
{Hu}, J.-F., {Cao}, X.-W., {Chen}, L., \& {You}, B. 2016, Research in Astronomy
  and Astrophysics, 16, 136, \dodoi{10.1088/1674-4527/16/9/136}

\bibitem[{{Isobe} {et~al.}(1990){Isobe}, {Feigelson}, {Akritas}, \&
  {Babu}}]{1990ApJ...364..104I}
{Isobe}, T., {Feigelson}, E.~D., {Akritas}, M.~G., \& {Babu}, G.~J. 1990, \apj,
  364, 104, \dodoi{10.1086/169390}

\bibitem[{{Jackson} \& {Browne}(1990)}]{1990Natur.343...43J}
{Jackson}, N., \& {Browne}, I.~W.~A. 1990, \nat, 343, 43,
  \dodoi{10.1038/343043a0}

\bibitem[{{Kruk} {et~al.}(2023){Kruk}, {Garc{\'\i}a-Mart{\'\i}n}, {Popescu},
  {Aussel}, {Dillmann}, {Perks}, {Lund}, {Mer{\'\i}n}, {Thomson}, {Karadag}, \&
  {McCaughrean}}]{2023NatAs...7..262K}
{Kruk}, S., {Garc{\'\i}a-Mart{\'\i}n}, P., {Popescu}, M., {et~al.} 2023, Nature
  Astronomy, 7, 262, \dodoi{10.1038/s41550-023-01903-3}

\bibitem[{{Laing} {et~al.}(1983){Laing}, {Riley}, \&
  {Longair}}]{1983MNRAS.204..151L}
{Laing}, R.~A., {Riley}, J.~M., \& {Longair}, M.~S. 1983, \mnras, 204, 151,
  \dodoi{10.1093/mnras/204.1.151}

\bibitem[{{Li} \& {Cao}(2019)}]{2019ApJ...886...92L}
{Li}, J., \& {Cao}, X. 2019, \apj, 886, 92, \dodoi{10.3847/1538-4357/ab4c36}

\bibitem[{{Li}(2014)}]{2014ApJ...788...71L}
{Li}, S.-L. 2014, \apj, 788, 71, \dodoi{10.1088/0004-637X/788/1/71}

\bibitem[{{Li} \& {Begelman}(2014)}]{2014ApJ...786....6L}
{Li}, S.-L., \& {Begelman}, M.~C. 2014, \apj, 786, 6,
  \dodoi{10.1088/0004-637X/786/1/6}

\bibitem[{{Li} \& {Gu}(2021)}]{2021A&A...654A.141L}
{Li}, S.-L., \& {Gu}, M. 2021, \aap, 654, A141,
  \dodoi{10.1051/0004-6361/202141301}

\bibitem[{{Li} {et~al.}(2022){Li}, {Zhou}, \& {Gu}}]{2022A&A...663L...4L}
{Li}, S.-L., {Zhou}, M., \& {Gu}, M. 2022, \aap, 663, L4,
  \dodoi{10.1051/0004-6361/202244105}

\bibitem[{{Liu} {et~al.}(2007){Liu}, {Taam}, {Meyer-Hofmeister}, \&
  {Meyer}}]{2007ApJ...671..695L}
{Liu}, B.~F., {Taam}, R.~E., {Meyer-Hofmeister}, E., \& {Meyer}, F. 2007, \apj,
  671, 695, \dodoi{10.1086/522619}

\bibitem[{{Lubow} {et~al.}(1994){Lubow}, {Papaloizou}, \&
  {Pringle}}]{1994MNRAS.267..235L}
{Lubow}, S.~H., {Papaloizou}, J.~C.~B., \& {Pringle}, J.~E. 1994, \mnras, 267,
  235, \dodoi{10.1093/mnras/267.2.235}

\bibitem[{{Lusso} \& {Risaliti}(2017)}]{2017A&A...602A..79L}
{Lusso}, E., \& {Risaliti}, G. 2017, \aap, 602, A79,
  \dodoi{10.1051/0004-6361/201630079}

\bibitem[{{Lusso} {et~al.}(2010){Lusso}, {Comastri}, {Vignali}, {Zamorani},
  {Brusa}, {Gilli}, {Iwasawa}, {Salvato}, {Civano}, {Elvis}, {Merloni},
  {Bongiorno}, {Trump}, {Koekemoer}, {Schinnerer}, {Le Floc'h}, {Cappelluti},
  {Jahnke}, {Sargent}, {Silverman}, {Mainieri}, {Fiore}, {Bolzonella}, {Le
  F{\`e}vre}, {Garilli}, {Iovino}, {Kneib}, {Lamareille}, {Lilly}, {Mignoli},
  {Scodeggio}, \& {Vergani}}]{2010A&A...512A..34L}
{Lusso}, E., {Comastri}, A., {Vignali}, C., {et~al.} 2010, \aap, 512, A34,
  \dodoi{10.1051/0004-6361/200913298}

\bibitem[{{Madrid} {et~al.}(2006){Madrid}, {Chiaberge}, {Floyd}, {Sparks},
  {Macchetto}, {Miley}, {Axon}, {Capetti}, {O'Dea}, {Baum}, {Perlman}, \&
  {Quillen}}]{2006ApJS..164..307M}
{Madrid}, J.~P., {Chiaberge}, M., {Floyd}, D., {et~al.} 2006, \apjs, 164, 307,
  \dodoi{10.1086/504480}

\bibitem[{{Marshall} {et~al.}(2018){Marshall}, {Avara}, \&
  {McKinney}}]{2018MNRAS.478.1837M}
{Marshall}, M.~D., {Avara}, M.~J., \& {McKinney}, J.~C. 2018, \mnras, 478,
  1837, \dodoi{10.1093/mnras/sty1184}

\bibitem[{{McKinney} {et~al.}(2012){McKinney}, {Tchekhovskoy}, \&
  {Blandford}}]{2012MNRAS.423.3083M}
{McKinney}, J.~C., {Tchekhovskoy}, A., \& {Blandford}, R.~D. 2012, \mnras, 423,
  3083, \dodoi{10.1111/j.1365-2966.2012.21074.x}

\bibitem[{{McLure} {et~al.}(2006){McLure}, {Jarvis}, {Targett}, {Dunlop}, \&
  {Best}}]{2006MNRAS.368.1395M}
{McLure}, R.~J., {Jarvis}, M.~J., {Targett}, T.~A., {Dunlop}, J.~S., \& {Best},
  P.~N. 2006, \mnras, 368, 1395, \dodoi{10.1111/j.1365-2966.2006.10228.x}

\bibitem[{{Merloni} \& {Heinz}(2007)}]{2007MNRAS.381..589M}
{Merloni}, A., \& {Heinz}, S. 2007, \mnras, 381, 589,
  \dodoi{10.1111/j.1365-2966.2007.12253.x}

\bibitem[{{Meyer} {et~al.}(2000){Meyer}, {Liu}, \&
  {Meyer-Hofmeister}}]{2000A&A...361..175M}
{Meyer}, F., {Liu}, B.~F., \& {Meyer-Hofmeister}, E. 2000, \aap, 361, 175,
  \dodoi{10.48550/arXiv.astro-ph/0007091}

\bibitem[{{Morales Teixeira} {et~al.}(2018){Morales Teixeira}, {Avara}, \&
  {McKinney}}]{2018MNRAS.480.3547M}
{Morales Teixeira}, D., {Avara}, M.~J., \& {McKinney}, J.~C. 2018, \mnras, 480,
  3547, \dodoi{10.1093/mnras/sty2044}

\bibitem[{{Narayan} {et~al.}(2003){Narayan}, {Igumenshchev}, \&
  {Abramowicz}}]{2003PASJ...55L..69N}
{Narayan}, R., {Igumenshchev}, I.~V., \& {Abramowicz}, M.~A. 2003, \pasj, 55,
  L69, \dodoi{10.1093/pasj/55.6.L69}

\bibitem[{{Netzer} {et~al.}(2006){Netzer}, {Mainieri}, {Rosati}, \&
  {Trakhtenbrot}}]{2006A&A...453..525N}
{Netzer}, H., {Mainieri}, V., {Rosati}, P., \& {Trakhtenbrot}, B. 2006, \aap,
  453, 525, \dodoi{10.1051/0004-6361:20054203}

\bibitem[{{Punsly}(2014)}]{2014ApJ...797L..33P}
{Punsly}, B. 2014, \apjl, 797, L33, \dodoi{10.1088/2041-8205/797/2/L33}

\bibitem[{{Punsly}(2015)}]{2015ApJ...806...47P}
---. 2015, \apj, 806, 47, \dodoi{10.1088/0004-637X/806/1/47}

\bibitem[{{Qiao} \& {Liu}(2018)}]{2018MNRAS.477..210Q}
{Qiao}, E., \& {Liu}, B.~F. 2018, \mnras, 477, 210,
  \dodoi{10.1093/mnras/sty652}

\bibitem[{{Runnoe} {et~al.}(2012){Runnoe}, {Brotherton}, \&
  {Shang}}]{2012MNRAS.422..478R}
{Runnoe}, J.~C., {Brotherton}, M.~S., \& {Shang}, Z. 2012, \mnras, 422, 478,
  \dodoi{10.1111/j.1365-2966.2012.20620.x}

\bibitem[{{Scepi} {et~al.}(2023){Scepi}, {Begelman}, \&
  {Dexter}}]{2023arXiv230210226S}
{Scepi}, N., {Begelman}, M.~C., \& {Dexter}, J. 2023, arXiv e-prints,
  arXiv:2302.10226, \dodoi{10.48550/arXiv.2302.10226}

\bibitem[{{Schlegel} {et~al.}(1998){Schlegel}, {Finkbeiner}, \&
  {Davis}}]{Schlegel_etal_1998}
{Schlegel}, D.~J., {Finkbeiner}, D.~P., \& {Davis}, M. 1998, \apj, 500, 525,
  \dodoi{10.1086/305772}

\bibitem[{{Shang} {et~al.}(2011){Shang}, {Brotherton}, {Wills}, {Wills},
  {Cales}, {Dale}, {Green}, {Runnoe}, {Nemmen}, {Gallagher}, {Ganguly},
  {Hines}, {Kelly}, {Kriss}, {Li}, {Tang}, \& {Xie}}]{2011ApJS..196....2S}
{Shang}, Z., {Brotherton}, M.~S., {Wills}, B.~J., {et~al.} 2011, \apjs, 196, 2,
  \dodoi{10.1088/0067-0049/196/1/2}

\bibitem[{{Shen} {et~al.}(2011){Shen}, {Richards}, {Strauss}, {Hall},
  {Schneider}, {Snedden}, {Bizyaev}, {Brewington}, {Malanushenko},
  {Malanushenko}, {Oravetz}, {Pan}, \& {Simmons}}]{2011ApJS..194...45S}
{Shen}, Y., {Richards}, G.~T., {Strauss}, M.~A., {et~al.} 2011, \apjs, 194, 45,
  \dodoi{10.1088/0067-0049/194/2/45}

\bibitem[{{Tchekhovskoy} {et~al.}(2011){Tchekhovskoy}, {Narayan}, \&
  {McKinney}}]{2011MNRAS.418L..79T}
{Tchekhovskoy}, A., {Narayan}, R., \& {McKinney}, J.~C. 2011, \mnras, 418, L79,
  \dodoi{10.1111/j.1745-3933.2011.01147.x}

\bibitem[{{Telfer} {et~al.}(2002){Telfer}, {Zheng}, {Kriss}, \&
  {Davidsen}}]{2002ApJ...565..773T}
{Telfer}, R.~C., {Zheng}, W., {Kriss}, G.~A., \& {Davidsen}, A.~F. 2002, \apj,
  565, 773, \dodoi{10.1086/324689}

\bibitem[{{Westhues} {et~al.}(2016){Westhues}, {Haas}, {Barthel}, {Wilkes},
  {Willner}, {Kuraszkiewicz}, {Podigachoski}, {Leipski}, {Meisenheimer},
  {Siebenmorgen}, \& {Chini}}]{2016AJ....151..120W}
{Westhues}, C., {Haas}, M., {Barthel}, P., {et~al.} 2016, \aj, 151, 120,
  \dodoi{10.3847/0004-6256/151/5/120}

\bibitem[{{White} {et~al.}(2019){White}, {Stone}, \&
  {Quataert}}]{2019ApJ...874..168W}
{White}, C.~J., {Stone}, J.~M., \& {Quataert}, E. 2019, \apj, 874, 168,
  \dodoi{10.3847/1538-4357/ab0c0c}

\bibitem[{{Wu} {et~al.}(2013){Wu}, {Brandt}, {Miller}, {Garmire}, {Schneider},
  \& {Vignali}}]{2013ApJ...763..109W}
{Wu}, J., {Brandt}, W.~N., {Miller}, B.~P., {et~al.} 2013, \apj, 763, 109,
  \dodoi{10.1088/0004-637X/763/2/109}

\bibitem[{{Xie} \& {Zdziarski}(2019)}]{2019ApJ...887..167X}
{Xie}, F.-G., \& {Zdziarski}, A.~A. 2019, \apj, 887, 167,
  \dodoi{10.3847/1538-4357/ab5848}

\bibitem[{{You} {et~al.}(2023){You}, {Cao}, {Yan}, {Hameury}, {Czerny}, {Wu},
  {Xia}, {Sikora}, {Zhang}, {Du}, \& {Zycki}}]{2023Sci...381..961Y}
{You}, B., {Cao}, X., {Yan}, Z., {et~al.} 2023, Science, 381, 961,
  \dodoi{10.1126/science.abo4504}

\bibitem[{{Yuan} \& {Narayan}(2014)}]{2014ARA&A..52..529Y}
{Yuan}, F., \& {Narayan}, R. 2014, \araa, 52, 529,
  \dodoi{10.1146/annurev-astro-082812-141003}

\bibitem[{{Yuan} {et~al.}(2022){Yuan}, {Wang}, \& {Yang}}]{2022ApJ...924..124Y}
{Yuan}, F., {Wang}, H., \& {Yang}, H. 2022, \apj, 924, 124,
  \dodoi{10.3847/1538-4357/ac4714}

\bibitem[{{Zamaninasab} {et~al.}(2014){Zamaninasab}, {Clausen-Brown},
  {Savolainen}, \& {Tchekhovskoy}}]{2014Natur.510..126Z}
{Zamaninasab}, M., {Clausen-Brown}, E., {Savolainen}, T., \& {Tchekhovskoy}, A.
  2014, \nat, 510, 126, \dodoi{10.1038/nature13399}

\bibitem[{{Zhao} {et~al.}(2023){Zhao}, {Yang}, {Xue}, \&
  {Li}}]{2023MNRAS.526..862Z}
{Zhao}, Y., {Yang}, X.-H., {Xue}, L., \& {Li}, S.-L. 2023, \mnras, 526, 862,
  \dodoi{10.1093/mnras/stad2816}

\bibitem[{{Zhou} \& {Gu}(2020)}]{2020ApJ...893...39Zhou}
{Zhou}, M., \& {Gu}, M. 2020, \apj, 893, 39, \dodoi{10.3847/1538-4357/ab7dca}

\bibitem[{{Zhu} {et~al.}(2020){Zhu}, {Brandt}, {Luo}, {Wu}, {Xue}, \&
  {Yang}}]{2020MNRAS.496..245Zhu}
{Zhu}, S.~F., {Brandt}, W.~N., {Luo}, B., {et~al.} 2020, \mnras, 496, 245,
  \dodoi{10.1093/mnras/staa1411}

\end{thebibliography}
\bibliographystyle{aasjournal}

\end{document}